\documentstyle[preprint,aps]{revtex}

\begin{document}

\title{
$E\otimes e$ Jahn-Teller Effect in  $C_{70}^{3-}$ Systems  
}
\author {Lin Tian,$^{1,2}$ Ya-Sha Yi,$^{2}$  
Chui-Lin Wang,$^{3}$ and Zhao-Bin Su$^{2}$}

\address{$^1$ Department of Physics, Peking University, Beijing 
100871, China} 
\address{$^2$ Institute of Theoretical Physics, Chinese Academy of 
Sciences, Beijing 100080, China}
\address{$^3$ China Center of Advanced Science and Technology
(World Laboratory),\\
Beijing 100080, China}
\date{January 26, 1997}
\maketitle

\begin{abstract} 
The electron-phonon interaction in $C_{70}$ anions
is studied by making use of a lattice relaxation approach.
We find there exists a Jahn-Teller effect in $C_{70}^{3-}$ system,
due to an extra electron being
doped to the double degenerate $E_{1}^{''}$ state.
As a result of this effect, the original  
$D_{5h}$ symmetry of the ground state becomes unstable, which
causes distortion of the lattice configuration. The only symmetry
maintained in the final state of the relaxation is the $x$-$y$ plane
reflection symmetry.
We further find that besides the Jahn-Teller active
$A_{1}^{'}$, $A_{2}^{'}$, $E_{2}^{'}$ modes, the
non-Jahn-Teller active $E_{1}^{'}$ vibrations also
contribute to the relaxation process, which come from the  
nonlinear effect and are two or three orders smaller
than those of the Jahn-Teller active modes.
We suggest that the $C_{70}^{3-}$ molecule
is a promising Berry Phase candidate in this
effective $E \otimes e$ Jahn-Teller system.
\end{abstract}
\newpage
 
In many physics systems, such as the low dimensional crystals and clusters,
both the electronic structure and the vibronic  
structure  are  affected by the electron-phonon coupling.
The $C_{70}$ molecule, with its $D_{5h}$ symmetry that includes double 
degenerate and nondegenerate irreducible representations, 
is  a perfect system to study the electron-phonon interaction in the 
electron-lattice systems. Since the discovery of $C_{70}$
about ten years ago\cite{Muller},
there are massive 
experiments exploring various properties of $C_{70}$, such as Raman 
spectra\cite{Beth,Raman}, photon luminescent and absorption 
spectra\cite{Wang,Hare,JETP}, {\sl et al}.
Many properties of the geometry
and electronic structure of $C_{70}$ have also been calculated by using
different methods such as quantum chemical calculation\cite{Negri},
first principle LDA\cite{Saito,Nakao}, vibrational 
potential method\cite{Suz}, mean field method\cite{Harris} 
and {\sl ab initial} Hartree-Fork method\cite{Scuseria,Reid}. 
But none of the above methods can obtain both the vibronic property 
and the electronic property self-consistently,
and all of them need huge computation.
\par
In this paper, we study the electron-phonon coupling in $C_{70}$ anions 
by using the simple and elegant Su-Schrieffer-Heeger (SSH) model. 
It is well known that the ground state of neutral $C_{70}$ molecule has 
$D_{5h}$ symmetry which is composed of 20 group elements including pentagon 
rotation and $x$-$y$ plane reflection. The $D_{5h}$ group has 
8 unequivalent irreducible representations (IR), $A_{1}^{'}$, $A_{2}^{'}$,
$A_{1}^{''}$, $A_{2}^{''}$, $E_{1}^{'}$, $E_{2}^{'}$, $E_{1}^{''}$ and
$E_{2}^{''}$. Both the electron levels and the phonon levels of
$C_{70}$ can be classified by these 8 IRs  
that are either single or two fold degenerate.
Generally, when the electron filling in molecules changes, 
the effective potential, which includes both the elastic energy 
between the lattice atoms and the electron-lattice coupling energy, 
is subject to modify accordingly. On the new potential surface,
the effective potential of the original lattice configuration deviates 
from the minimum points and the system becomes unstable, that will result in  
a lattice relaxation process until a new 
stable point finally being reached. During this process,
the lattice symmetry may be broken depending on electionic dopping.
If an additional electron is dopped to a nondegenerate level, 
only the total symmetric modes participate in the lattice relaxation
according to the Jahn-Teller theorem. 
The original symmetry of the lattice will not be broken by these modes
and the degenerate electron levels will not split.  
However, when the electron is filled to a degenerate level,
the Jahn-Teller active
modes will include non-total symmetric vibrations
besides the total symmetric modes.
Lattice displacements along the directions of these non-total symmetric
modes will reduce both the lattice symmetry and the degeneracy
of the electron levels.
In neutral $C_{70}$, the highest occupied molecular orbital
(HOMO) and the lowest unoccupied molecular orbital (LUMO)
are both nondegenerate, with $A_{2}^{''}$ and $A_{1}^{''}$ symmetry 
respectively, while the next higher level (the 37th, 38th level) above the 
Fermi surface is double degenerate. 
The salient difference between  $C_{70}^{-}$ or $C_{70}^{2-}$
(lighter doping) and $C_{70}^{3-}$ (heavier doping) is that the former
case only involves eletronic filling in a non degenerate level whereas
the latter involves in a degenerate one.
In the cases of $C_{70}^{-}$ and $C_{70}^{2-}$ clusters where the
extra electrons are doped to the $A_{1}^{''}$ LUMO level, 
the $D_{5h}$ symmetry is maintained in the final state, since the lattice displacement 
from the original configuration contents only  $A_{1}^{'}$ vibration modes
which are totally symmetric. 
In the case of $C_{70}^{3-}$, the third electron
is filled to the next higher level ($E_{1}^{''}$ representation), 
after the LUMO is fully occupied by the other two extra electrons.  
The Jahn-Teller active modes in this case include $A_{1}^{'}$, 
$A_{2}^{'}$ and  $E_{2}^{'}$ vibrations. As a result, the original 
$D_{5h}$ symmetry is broken after relaxation and all the electron 
levels become nondegenerate. From further analysis (see below),
we find that all the participating modes in this process will maintain the 
the $x$-$y$ plane reflection symmetry, 
which is the only symmetry left after the lattice relaxation.
It is worth to point out that other non-Jahn-Teller modes
besides the Jahn-Teller active modes may also make contribution
once the lattice relaxation evolves.
This is due to the non-linear effect, which is involved in
coupling between the vibration modes in the effective potential.
The nonzero contribution of the non-Jahn-Teller active modes
is proportional to the displacement of the Jahn-Teller active mode, so that
the displacements along these non-Jahn-Teller active modes may be 
substantially smaller than that of the Jahn-Teller active modes.
We find that in$C_{70}^{3-}$  anion there only exists one kind of
non-Jahn-Teller active modes, $E_{1}^{'}$ modes, whose amplitudes
are two or three orders smaller than that of 
the $A_{1}^{'}$, $A_{2}^{'}$, and $E_{2}^{'}$ vibrations. From the above
analysis, it is clear that the $C_{70}^{3-}$ system reduces to
an effective $E\otimes e$ Jahn-Teller model\cite{Bersuker,ham} which carries 
the Berry Phase, where $E$ and $e$ are 
the group representations of the electron and vibration modes respectively. 
We suggest that the $C_{70}^{3-}$ system is a promising candidate to verify  
the Berry Phase and the corresponding selection rules featured by the 
$E\otimes e$ Jahn-Teller model. 
As it is relatively easy to charge $C_{70}$ in laboratories,
our results may be observed and verified by further experiments.

\par
The SSH model  has the following form\cite{You}:
\par
\begin{equation}\begin{array}{cl}
H=-\displaystyle\sum_{\langle i,j\rangle,s} 
(t_{0}-\alpha(l_{ij}-l_{1}))(c^{\dagger}_{is}c_{js}+h.c)\\
+\displaystyle\frac{1}{2}K_{1} \sum_{\langle i,j\rangle}
(l_{ij}-l_{1})^{2}\\
+\displaystyle\frac{1}{2}K_2 \sum_{i^{'}}d\theta_{i,5}^{2}
+\frac{1}{2}K_3\sum_{i^{'}}(d\theta_{i,6,1}^{2}
+d\theta_{i,6,2}^{2})\\
+\displaystyle\frac{1}{2}K_4 \sum_{i^{''}}d\theta_{i,6,3}^{2}
+\frac{1}{2}K_5\sum_{i^{''}}(d\theta_{i,6,1}^{2}
+d\theta_{i,6,1}^{2})\\
\end{array}
\end{equation}
where $t_{0}$ is the hopping integral 
between  nearest neighbour carbon atoms;
$\alpha$ is the e-ph coupling constant; and $l_{ij}$ is 
the bond length between different atoms with $l_{1}$ being
the bond length with all
the bond stretching springs relaxing.
$c_{is}^{\dagger}$ ($c_{is}$) are the 
electron creation (annihilation) operators of the $\pi$ electron on site $i$
with spin index $s$.  $d\theta_{i,5}$, $d\theta_{i,6,1}$, 
$d\theta_{i,6,2}$, $d\theta_{i,6,3}$  are bond angle deviations 
from the original angle,  
$108^{\circ}$,$120^{\circ}$,$120^{\circ}$,$120^{\circ}$ respectively, 
$e.~g. $ \( d\theta_{i,5}=\theta_{i,5}-108^{\circ} \), {\sl etc}. $K_{i}$ 
\( (i=1\) to $5$)  are  the elastic constants for 
different kinds of lattice vibrations. 
The primes above the summation in the
$K_{2}$ and $K_{3}$ terms indicate  exclusion of the middle 
ten atoms near the equator, while the double primes in the $K_{4}$
and $K_{5}$ terms denote
that the summation includes only the equatorial ten atoms because 
the  geometry condition for them are different from the others. 
The first term in this formula is the tight binding electron Hamiltonian
describing $\pi$ electrons hopping between the
nearest neighbour sites.
The $K_{i}$  terms are the elastic energies of
the lattice atoms in the form of classical springs. 
The kinetic energies of atoms are not included in the Hamiltonian in the
adiabatic approximation. The springs of the angular terms
are introduced to represent the degrees of  freedom that can not be
represented by pure bond-length terms. 
The $\alpha$ term inside the tight-binding electron Hamiltonian represents
the linear electron-phonon interaction which is valid when the lattice 
displacements are small.
We use the following set of
parameters: \(\alpha=6.20 \) eV/{\AA}, \(t_{0}= 
2.50\) eV, \(K_{1}=42\) eV/$\AA^{2}$, \(K_{2}=8.0\) eV/$rad^{2}$,
\(K_{3}=7.0\) eV/$rad^{2}$, \(K_{4}=6.5\) eV/$rad^{2}$,
\(K_{5}=6.0\) eV/$rad^{2}$. Our results fit well with experimental
observable data which show that these parameters are reasonable.
The coulomb interaction is not considered in this 
model as the electron-electron interaction contributes little to the lattice 
relaxation. This conclusion is came from 
another work\cite{tian}, where the coulomb interaction is 
analysized and shows no important effects on the Jahn-Teller system. 
We solve this model in a framework of
the Bogoliubov-De Gennes(BdeG) formalism and get the stable
point in the adiabatic potential surface self-consistently. Both the lattice 
configuration and the electron state  are derived within simple 
calculation. The advantage of this model is that 
we can get the electronic and the vibronic properties together. 
Comparing it with more complex and sophisticated 
methods, such as LDA, the SSH model only has a few parameters but 
grasps the essential physics. 
Our calculations show that it is a suitable model to deal with the fullerene 
family.
\par
We apply the dynamical method\cite{You} to approach the stable
point gradually on the effective potential surface. 
Starting from an initial configuration, we calculate the
electron levels  and the effective potential $V_{\rm eff}$, 
derive the force on each atom from the $V_{\rm eff}$
that is exerted by both the electrons and other atoms, 
and get lattice displacements that are caused by
these forces, then start a new round by using
new configuration.
This dynamic process is repeated until 
the stable point (the minimum) on the potential surface being reached.
The following equations are necessary:

\begin{equation}\begin{array}{cl}
\displaystyle
F_{i\sigma}=-\frac{dV_{\rm eff}}{dx_{i\sigma}}\\[5mm]
\displaystyle
v_{i\sigma}=\frac{dx_{i\sigma}}{dt}\\[5mm]
\end {array}
\end {equation}
where $V_{\rm eff}$ is the effective lattice potential that includes the 
elastic energy between the lattice atoms and the electron-lattice interaction
energy that also is function of the lattice coordinates.
$F_{i\sigma}$ is a force component 
in direction $\sigma$ on atom $i$ 
that is derived from 
the effective lattice potential.

\par
We can also derive the phonon modes by this simple model.
Making use of the final configuration of the relaxation, we can derive
the vibration frequencies and vectors 
by adding the kinetic parts to the lattice Hamiltonian and applying the 
elastic coefficient matrix method.
The following equations are used:

\begin{equation}\begin{array}{cl}
H_{\rm eff}=\displaystyle\frac{1}{2} m\sum_{i,s}(\frac{dx_{i,s}}{dt})^{2} 
+\frac{1}{2}K_{i,s ;j,s^{'}}
x_{i,s}x_{j,s^{'}}\\[5mm]
K_{i,s; j,s^{'}}=
 \displaystyle\frac{\partial^{2}V_{\rm eff}}{\partial x_{i,s}
\partial x_{j,s^{'}}}\\[5mm]
V_{\rm eff}=V_{elas} + V_{e-ph}\\[5mm]
\end{array}
\end{equation}
where
$x_{i,s}$ is a small displacement of atom $i$ from its equilibrium
position in the $s$
direction, $K_{i, s; j, s^{\prime}}$ is an element
of the elastic coefficient matrix. 
The first term of the Hamiltonian
is the kinetic energy of the lattice atoms
and the second term is the effective potential energy  
related to the equilibrium position of the lattice. 
$V_{\rm eff}$ is composed of two parts:
the elastic energy of the lattice $V_{elas}$ and the e-ph 
coupling energy $V_{e-ph}$, that is 
derived by the second order 
perturbation\cite{You} from the e-ph coupling Hamiltonian and by
omission of the higher order parts.
Using Eq. (3), we get the elastic coefficient
matrix and find the eigenvalues and eigenvectors, which are just the
phonon modes. The group representation of these vibration modes
can be determined by the projection
method that uses characters of $D_{5h}$ group elements.
\par
The first results we get are the lattice structures of the neutral 
$C_{70}$ and $C_{70}$ anions, all of that have the shape of prolate.
We list the bond lengths and angles of these clusters in Table (1).
Having $D_{5h}$ symmetry in $C_{70}$, $C_{70}^{-}$,
and $C_{70}^{2-}$, there are 3 kinds of bond lengths in these clusters.
The short bosigeleheaeef thf thfa(ed.ed.ea de$ I$ I$ent iavavaoms ipaˆf thf450\a de$
besides these two kinds of ava there is another type of ava whose 
length is aboueef.420\AA, and its value is between the long and short ava.
The middle length bonds have the same character as the ava in the 
Graphite. The main difference among these $C_{70}$ clustero that
the dopped clusters has longer bond length
in the neighbourhood of the equator, whose structure is looser than the
neutral one.
The electron density near the equator plane is larger,
hence the electron screening of the lattice elastic interaction is stronger.
This phenomena can explain the bond length's difference.
As a result, the skeleton of the $C_{70}^{2-}$ is looser
than that of the  $C_{70}^{-}$ molecule.
It seems that the central ten atoms are inserted equatorially in the $C_{60}$
ball like a loose chain withoue
dimerization. This point can also be seen in the bond angles which are almost
the same as that of $C_{60}$ for all atoms except the equatorial ones, that
have the largest deviation of $2.0^{\circ}$ from $120^{\circ}$ in the 
hexagon. 
The small deviation of the angles shows that the enlongation of the
$C_{70}$ molecule from the ball shape is very small.
In $C_{70}^{3-}$, the pentagon symmetry $C_5$ is broken, and  the bond 
angles and bond lengths in the same layer of the prolate are not the same, 
but the overall structure is also a prolate with very small deviation.
\par
In neutral $C_{70}$, as the lattice symmetry of $D_{5h}$ is relatively lower
than the $I_h$ symmetry of $C_{60}$, 
there are only double degenerate and non-degenerate electron levels  
which locate close to each other.
The electrons fill in the lower 35 levels and all
the anti-bond states are empty. The total symmetry of this many electron 
state is $A_{1}^{'}$.
The gap between HOMO and LUMO is 2.14 eV, which 
agrees well with the experimental result\cite{Wang}. 
This certifies that the set of parameters we choose is reasonable since the gap 
value varies sensitively with the parameters used in the calculation.
The group representations of HOMO and LUMO are 
$A_{2}^{''}$ and $A_{1}^{''}$  respectively,  
both of them being non-degenerate states. Hence, if extra electrons
are charged to the LUMO, which is non degenerate, which is the case of 
$C_{70}^{-}$ or $C_{70}^{2-}$, the $D_{5h}$ symmetry will not be lowered and
the Jahn-Teller effect will not occur. Another such example is the lowest energy
exciton with one electron being excited from the HOMO to the LUMO. This 
exciton is an $A_{2}^{'}$ state which also is non-degenerate\cite{tian}.
In these cases, the lattice displacement 
between the final state
and the initial state before relaxation 
contents only the total symmetric
$A_{1}^{'}$ vibrations which will keep the $D_{5h}$ symmetry, as is 
listed in Table (3.a). 
But this is not the case in the $C_{70}^{3-}$, 
as will be discussed in detail later.
The total energy in the new configuration is lower than that 
of the initial one, hence the new state is more stable. 
In $C_{70}^{-}$, the gap drops to
f thf91 eV after the lattice relaxation; 
in $C_{70}^{2-}$, when the LUMO is filled up, 
the gap drops further to f thf69 eV in the stable state, that indicates 
the gap is reduced upon doping. 
>From the energy spectrum, we find that the LUMO drops greatly,
and the HOMO level and all the other states 
almost do not change upon dopping.
Thio the polaron effect which results from the electron-phonon 
coupling.
The electron wavefunction is localized near the plane next to the equator
as it has been shown by the electron density on atoms.
\par
In an electron-lattice interaction system,
the electron filling can affect the atomic 
configuration, and in return, the lattice configuration will determine the 
electron wavefunctions and levels also. The difference between the electron 
densities in various electron fillings manifests this mutual interaction. 
When the electron density in a site is
denser, the electron-lattice interaction will
screen more of the elastic force acting on the nearby atoms. Then, the bond 
length related to the atoms will become more relaxed and becomes longer. 
In Table (2), our results for the electron densities in 
five kinds of points (which were certified by the NMR 
experiments\cite{Muller}) in different kinds of clusters are listed. 
The density differences between different
points  are small since $C_{70}$ is 
approximately spherical with the height along the pentagon axis being
a little longer than the diameter of the equator plane. 
In the doped clusters, the electron densities on atoms 
near the equtorial plane and around the two tops 
do not change much from that of the neutral cluster. 
The additional electron density due to dopping
distributes mainly on the plane that is next to the equator. 
It agrees with the bond length distribution. 
Upon doping, ava $6$ and $4$ become longer due to the screening 
of the elastic interaction.
\par
In $C_{70}^{3-}$ anion, when the third extra electron is doped to the 
level immediately higher than LUMO, which
is a double degenerate $E_{1}^{''}$ level, 
the $D_{5h}$ symmetry of the initial lattice configuration
is destroyed during relaxation,  
and all the electron levels become
non-degenerate, due to the static Jahn-Teller effect.  
The motion of the lattice coordinate can be discussed on the 
effective potential surface in the configuration space.
When the electron filling changes, the lattice configuration jumps from 
the initial potential surface to another surface on which the lattice 
configuration is not at the minimum point of the potential. 
Then, the initial configuration becomes unstable
and the cluster begins the lattice relaxation process. At the point
when the relaxation just starts, 
the vibration modes that couples to the lattice motion 
include only the Jahn-Teller active modes. At this point, the 
derivatives of effective potential in 
the Jahn-Teller active vibration directions are not zero, 
while the derivatives along all other directions are zero. 
\par
Phonon modes are Jahn-Teller active when they satisfy the 
following equation:
\begin{equation}
 \displaystyle\langle e| \frac{\partial{H}}{\partial{Q_{\nu}}} | e \rangle \ne 0
\end {equation}
where $Q_{\nu}$ is the displacement along phonon mode $\nu$, 
and $|e \rangle$ is  the 
electron state that depends on the lattice displacement. 
Using symbols of group theory, Eq.(4) can be interpreted as follows:
\begin{equation}
 G^{e} \otimes \overline{G^{e} }\otimes 
\overline{ G^{ph} } \supset A_{1}^{'}
\end {equation}
where $G^{e}$ is the group representation of 
multielectron state with $\overline{G^{e} }$ being its
complex conjugate counterpart. Here, \( G^{e}=\overline{ G^{e}}\)
since the representation is real. 
$\overline{G^{ph}}$ is the complex conjugate representation 
of phonon mode. In the $C_{70}^{3-}$, 
the above equation is equivalent to:
\begin{equation}
E_{1}^{''} \otimes E_{1}^{''} \supset A_{1}^{'} 
\oplus A_{2}^{'} \oplus E_{2}^{'}
\end{equation}
  From this equation, not only the total symmetry $A_{1}^{'}$ vibrations
contribute to the dynamical process in the beginning, but $E_{2}^{'}$ and 
$A_{2}^{'}$ 
modes may also participate in the process. These three kinds of modes 
are the Jahn-Teller active modes that will
break the lattice symmetry. From the first step on,
the lattice will reach to the new minimum point on the new
potential surface after a relaxation period.
All the participating modes have $x$-$y$ plane reflection symmetry. So, 
only the $x$-$y$ plane 
reflection symmetry is left unperturbed while the pentagon 
rotation symmetry is broken, and 
the degenerate electron levels split into single levels. 
\par
All the above analytical results
can be checked numerically by using the dynamical method. 
Within the simple SSH model, we can get all the quantities, 
the lattice configuration of the neutral $C_{70}$ and the $C_{70}^{3-}$, 
and the phonon canonical vectors.
The lattice displacement during the relaxation process is just the
difference between 
the lattice coordinates of the initial and the final state.
We  project this displacement to 
the normal vector of the phonon modes to get amplitude $Q_{\nu}$
in $\nu$ direction.
Let $x_{i,\sigma}^{(0)}$ be the lattice position of the neutral cluster,
and $x_{i,\sigma}^{(3)}$ be the lattice 
position of the $C_{70}^{3-}$ anion, then,

\begin{equation}
Q^{\nu}=\sum_{i,\sigma}(x_{i \sigma}^{(3)}-x_{i \sigma}^{(0)})\xi_
{i,\sigma}^{\nu}
\end {equation}
where $\xi_{i,\sigma}^{\nu}$ is the phonon vibration 
vector. Using this method, we derive the 
contributions from  different phonon modes.
In our numerical results, the modes that 
give non-zero contributions are  $A_{1}^{'}$, $E_{2}^{'}$, and $E_{1}^{'}$
among which $A_{1}^{'}$ and $E_{2}^{'}$ are  Jahn-Teller active modes.
All the amplitudes in $E_{1}^{'}$ are two or three orders smaller than 
those of the $A_{1}^{'}$ and $E_{2}^{'}$ modes.
The absence of the Jahn-Teller $A_{2}^{'}$ modes
does not contradict with the Jahn-Teller theorem. 
After the first step in the relaxation process, 
the symmetry of the cluster is decreased, 
other phonon modes may join the process too. 
This can be explained by the effective potential approach. 
We can write the effective potential in Taylor expansion
and select the initial position as the zero point.
The initial force on the 
lattice atoms can  be projected only to the Jahn-Teller direction.
The force at the zero point only results from the 
first order terms of the effective potential since all the higher order 
terms at this point are zero. 
But, if the displacement from the initial point 
is not zero, the higher order terms may also make contribution. 
There are coupled terms between different lattice vibrations, 
hence, non-Jahn-Teller modes may participate the relaxation. 
Since the forces from the second order terms 
are proportional  to  the lattice displacement, it will be several orders 
smaller because the lattice distortion is not large. 
In $C_{70}^{3-}$, we can find  
that the modes that couple to the Jahn-Teller modes in the second order 
terms of the effective potential are only the $E_{1}^{'}$ modes. Our 
calculation certifies this point. 
And the $E_{1}^{'}$ displacements are reasonably small.
In Table (3.b), we list the amplitudes of those modes that 
give non-zero contributions as well as the frequency of those modes. The 
unit of the frequency is $cm^{-1}$. In table (3.a), we also list
lattice displacements  of the $C_{70}^{-}$, $C_{70}^{2-}$
and $C_{70}$ exciton, which are nonzero only in the $A_{1}^{'}$
directions, since the corresponding single electron states are nondegenerate.
\par
Since the third doping-electron in $E_{1}^{''}$ state couples strongly to 
$E_{2}^{'}$ state, it turns out the $C_{70}^{3-}$ system can be
represented by an effective $E\otimes e$ Jahn-Teller model
that has been thoroughly studied for several decades\cite{Bersuker,ham}.
It is well proved that there exists Berry Phase in the $E \otimes e$
effective system. The consequent selection rules
found in this system are intriguing and
have been observed by many experiments\cite{ham}.
Hence we may further suggest the possibility of the presence of the Berry
Phase, $i.~e.$, the sign change in the electronic wave functions under a
$2\pi$ rotation in the space of the vibrational coordinates,
in $C_{70}^{3-}$ system which might be
the first polymer sample to manifest the fascinating $E \otimes e$ model.
We wish our prediction can be observed experimentally within the
reach of the current technology. 
\par
In conclusion, we have carefully studied the electron-phonon
interaction in the $C_{70}$ molecule.
In particular, the static Jahn-teller effect in the $C_{70}^{3-}$
is discussed. The lattice displacement from the neutral $C_{70}$ to the
$C_{70}^{3-}$ anion is projected to the canonical vibration vector space..
The nonzero contributions among these projections are the Jahn-Teller 
active $A_{1}^{'}$ and $E_{2}^{'}$ modes, as well as
the non-Jahn-Teller active $E_{1}^{'}$ modes that
couple to the Jahn-Teller active modes through the second order
non-linear terms in the effective potential.
The contributions of $E_{1}^{'}$ modes are comparably
smaller than those of the Jahn-teller active modes since the former
depends on the lattice distortions that are very small.
We find that $C_{70}^{3-}$ is an effective $E\otimes e$ e-ph
system, and the Berry Phase might be presented in such a system.
Our results show that the SSH model is a convenient and self-consistent
model to study the electron-lattice interaction in the fullerene family.
\par
\acknowledgements
This work was partly
supported by the National Natural Science Foundation of China.

\newpage

\small{
\begin{description}
\item{ {\bf Table 1 (a)}} Listed below are
bond lengths of $C_{70}$ and its anions. The results of $C_{70}^{3-}$
are not listed due to the broken symmetry.
\end{description} 
\begin{tabular}{|l|l|l|l|l|l|l|l|l|} \hline\hline
     { }      &ˆf    &ˆ 2    & 3    & 4 &ˆ5  &ˆ6  & 7   &ˆ8    \\ \hline
 $C_{70}$        &f thf447 &ˆfelectron dea97 &f thf446 &ˆf thfa(e3  
&ˆf.453 &ˆf thf421 &ˆf.418 &ˆf.448 \\ 
 $C_{70}^{-}$     &f thf448 &ˆfea(e8 &f thf446 &ˆf.403  
&ˆf.441 &ˆfe431 &ˆf thf420 &ˆf.445 \\
 $ C_{70}^{2-}$    &f thf448 &ˆf.a(e8 &f.446 &ˆfe413  
&ˆfe429 &ˆf.443 &ˆf.420 &ˆf.443 \\ \hline\hline
\end {tabular} }

\small{ \begin{description}
\item{ {\bf Table 1 (b)}} The calculated bond angles from a through l of 
		$C_{70}$ and $C_{70}^{\pm}$ together with experimental 
		results are listed. The $\dagger$ has the same meaning 
		as that in table(1).
\end{description} 
\begin{tabular}{|c|c|c|c|c|c|c|c|c|c|c|c|c|} \hline\hline
 {   }	          &ˆ a    &   b   &   c   &   d  
&ˆ   e   &ˆ f    &  g    &   h   &   i   &  j    &   k
&ˆ l     \\ \hline
 $ C_{70}$        &ˆf08.0 &ˆf19.9 &ˆf20.0 &ˆf07.7
&ˆf20.1  &ˆf07.9 &ˆ120.1 &ˆf20.3 &ˆf19.1 &ˆf08.3 &ˆ118.0
&ˆf19.6   \\ 
 $ C_{70}^{-}$    &ˆf08.0 &ˆf20.0 &ˆf20.0 &ˆf07.6
&ˆ119.9  & 108.0 &ˆ120.0 &ˆ120.5 &ˆf19.0 &ˆ108.2 &ˆf18.1
&ˆ119.5   \\ 
 $ C_{70}^{2-}$    & 108.0 &ˆ120.1 &ˆf20.0 &ˆ107.7
&ˆf19.8  &ˆf08.0 &ˆf20.0 &ˆf20.5 &ˆf19.0 &ˆf08.2 &ˆf18.1
&ˆ119.5   \\  \hline\hline
\end {tabular}
}

\small{
\begin{description}
\item{ {\bf Table 2}} Electronic densities on 5 kinds of positions 
		from A to E of $C_{70}$ and $C_{70}^{\pm}$ 
		are shown.The $\dagger$ has the same meaning 
		as that in table(1).
\end{description} 
\begin{tabular}{|c|c|c|c|c|c|} \hline\hline
 {  }		  &  A    & B     &   C   &   D   &  E    \\ \hline
 $  C_{70}$       & 0.991 &ˆf.013 &ˆ0.996 &ˆf.012 &ˆ0.980 \\
 $  C_{70}^{-}$   &ˆ0.988 &ˆf.015 &ˆf.021 &ˆf.038 &ˆ0.980 \\ 
 $  C_{70}^{2-}$   &ˆ0.986 &ˆf thf016 &ˆf.046 &ˆf.064 &ˆ0.979\\ \hline
\hline
\end {tabular}}

\small{
\begin{description}
\item{ {\bf Table 3 (a)}} Lattice relaxations between ground state and the
		lowest exciton state, $C_{70}$ anions are projected to different phonon
		modes.The results listed in this table are projections
		to $A_{1}^{'}$ modes. The contribution of all
			other modes are zero.
\end{description} 
\begin{tabular}{|l |c c c c c c c c c c c c|} \hline\hline
&ˆf3      &      23   &ˆ    38  &ˆ 53     &    59   &    93   
&ˆ125    &   138     &     142 &  154    &  178    &ˆ 210   \\ \hline   
 $C_{70}^{-}$ &-0.058  &    0.020  &ˆ -0.008 &ˆ0.002   &  0.008  &ˆ0.014 
& -0.022 &ˆ  0.020   &ˆ 0.009 &ˆ0.004  &ˆ 0.006  &ˆ 0.014 \\ \hline   
\hline
 $C_{70}^{2-}$ &-0.159  &ˆ   0.048  &  -0.017 &ˆ0.005   &ˆ 0.017  &ˆ0.028 
& -0.044 &   0.040   &ˆ 0.017 &ˆ0.008 &ˆ 0.012  &ˆ 0.028 \\ \hline   
\hline
 Exciton &-0.138  &ˆ   0.035  &  -0.022 &ˆ0.015   &  0.015  &ˆ0.026 
& -0.041 &ˆ  0.036   &ˆ -0.002 & -0.002  &  0.009  &  0.025 \\ \hline   
\hline
\end {tabular} }
\small{
\begin{description}
\item{ {\bf Table 3 (b)}} Lattice relaxations between ground state and the
		lowest exciton state, $C_{70}^{3-}$ are projected to different
		phonon modes.
\end{description} 
\begin{tabular}{| l | l | l | l | l | l | l | l | } \hline\hline
$A_{1}^{'}$  (/cm) &ˆ&ˆ $E_{2}^{'}$  (/cm) & &ˆ
&ˆ $E_{1}^{'}$  (/cm) &ˆ&ˆ\\ \hline
229  & -0.178 &ˆ187 &ˆ-0.040 &ˆ-0.021 &ˆ240 &ˆ-1.5*10$^{-4}$ &ˆ
-1.2*10$^{-4}$ \\
302  &ˆ0.050  & 219 &ˆ0.010  &ˆ-0.018 &ˆ275 &ˆ-1.1*10$^{-5}$ &
-7.5*10$^{-8}$ \\
380  & -0.007 & 317 & -0.016 &ˆ-0.009 &ˆ315 &ˆ-f thf1*10$^{-4}$ &
f thf6*10$^{-4}$ \\
452  & 0.012  & 371 &ˆ-0.002 & 0.004  & 347 &ˆ-2.0*10$^{-5}$ &
2.9*10$^{-5}$ \\
479  & 0.019  & 413 &ˆ-0.001 &ˆ6.2*10$^{-4}$&ˆ422 & 
-f thf1*10$^{-4}$ & -9.9*10$^{-5}$ \\
687  &ˆ0.031  & 477 & -0.005 &ˆ0.002  & 4a(e &ˆ-3.4*10$^{-5}$ &
2.9*10$^{-5}$ \\
f050 &ˆ-0.042 & 494 & -0.003 &ˆ-0.006 &ˆ524 &ˆ3.5*10$^{-5}$  &
-3.1*10$^{-5}$ \\
1232 &ˆ0.043  & 555 &ˆ-0.003 &ˆ0.006  &ˆ538 &ˆ-4.0*10$^{-5}$ &
-5.5*10$^{-5}$ \\
1299 &ˆ0.033  &ˆ611 &ˆ-0.002 & 0.003  & 565 &ˆ-f thf9*10$^{-5}$ &
-1.7*10$^{-5}$ \\
1357 & 0.022  & 711 &ˆ0.007  & -0.004 & 686 &ˆ5electron dea*10$^{-5}$  &
7.0*10$^{-5}$ \\
1512 & 0.009  & 736 &ˆ-0.006 &ˆ-0.004 & 875 &ˆ-5e3*10$^{-5}$ &
-6.4*10$^{-5}$ \\
1724 &ˆ0.030  & 846 &ˆ4.6*10$^{-4}$ & 2.5*10$^{-4}$ & 933 &
-5e6*10$^{-5}$ &ˆ-6.9*10$^{-5}$ \\
     &        &ˆ9a(e &ˆ0.007  & -0.004 &ˆf111 &ˆf.4*10$^{-4}$ &
f.0*10$^{-4}$ \\
     &        &ˆf024&ˆ-0.002 &ˆ0.001  &ˆf202 &ˆ9.2*10$^{-5}$ &
-1.1*10$^{-4}$ \\
     &        & 1235&ˆ0.001  & -0.001 &ˆf340 &ˆ-8.8*10$^{-5}$ &
-f thf1*10$^{-4}$ \\
     &        & 1325& 0.003  &ˆ-0.003 &ˆ1371 &ˆf.1*10$^{-4}$ &
-8.4*10$^{-5}$ \\
     &ˆ       & 1377&ˆ-0.004 & 0.003  & 1419 &ˆ-6.7*10$^{-5}$ &
8.7*10$^{-5}$ \\
     &        & 1593&ˆ6.0*10$^{-4}$ & 0.002 &ˆf477 &ˆ-4.3*10$^{-5}$ &
2.6*10$^{-5}$ \\
     &        & 1623& -0.005 &ˆ0.003  & 1504 &ˆ2.1*10$^{-5}$ &
2.6*10$^{-5}$ \\
     &        & 1708&ˆ-5e0*10$^{-4}$ & 0.030 &ˆ1605 &ˆ3.0*10$^{-5}$ &
-3.9*10$^{-5}$ \\
     &        & 1738&ˆ0.010  & 0.008  & 1680 &ˆ1.1*10$^{-5}$ &
-5.0*10$^{-4}$ \\ 
     &        &     &ˆ       &        &ˆf728 &ˆ-2.4*10$^{-5}$ &
f thf5*10$^{-5}$ \\ \hline
\hline
\end {tabular} }

\end{document}